\documentclass[aps,floatfix,footinbib,amsmath,amssymb,pra,twocolumn,superscriptaddress,longbibliography]{revtex4-2}

\usepackage{graphicx}
\usepackage{epstopdf}

\usepackage[applemac]{inputenc}
\usepackage[T1]{fontenc}
\usepackage{lmodern}
\usepackage[english]{babel}
\usepackage{ae}
\usepackage{units}
\usepackage{color}
\usepackage{url}

\usepackage{dcolumn}
\usepackage{bm}

\usepackage{url}
\usepackage[colorlinks]{hyperref}
\hypersetup{%
 plainpages=true,
 breaklinks=true,       
 hypertexnames=false,  
 pageanchor=true,
 colorlinks=true,
 linkcolor={blue},
 citecolor={blue},
 urlcolor={blue},
 anchorcolor={black}
}

\usepackage{mleftright} 

\newcommand{\ket}[1]{\left|#1\right\rangle}
\newcommand{\bra}[1]{\left\langle#1\right|}
\newcommand{\be}{\begin{equation}}
\newcommand{\ee}{\end{equation}}
\newcommand{\bea}{\begin{eqnarray}}
\newcommand{\eea}{\end{eqnarray}}

\newcommand{\figref}[1]{\mbox{Fig.~\ref{#1}}}

\newcommand{\appref}[1]{\mbox{Appendix~\ref{#1}}}
\renewcommand{\eqref}[1]{\mbox{Eq.~(\ref{#1})}}

\newcommand{\figpanel}[2]{Fig.~\hyperref[#1]{\ref*{#1}(#2)}}
\newcommand{\figpanels}[3]{Fig.~\hyperref[#1]{\ref*{#1}(#2)-(#3)}}
\newcommand{\figpanelNoPrefix}[2]{\hyperref[#1]{\ref*{#1}(#2)}}

\begin{document}

\title{Deterministic one-way logic gates on a cloud quantum computer}

\author{Zhi-Peng Yang}
\affiliation{Ministry of Education Key Laboratory for Nonequilibrium Synthesis and Modulation of Condensed Matter, Shaanxi Province Key Laboratory of Quantum Information and Quantum Optoelectronic Devices, School of Physics, Xi'an Jiaotong University, Xi'an, Shannxi, 710049, China}
\affiliation{Theoretical Quantum Physics Laboratory, RIKEN Cluster for Pioneering Research, Wako-shi, Saitama 351-0198, Japan}

\author{Huan-Yu Ku}
\email{huan_yu@phys.ncku.edu.tw}
\affiliation{Theoretical Quantum Physics Laboratory, RIKEN Cluster for Pioneering Research, Wako-shi, Saitama 351-0198, Japan}
\affiliation{Department of Physics and Center for Quantum Frontiers of Research \& Technology (QFort), National Cheng Kung University, Tainan 701, Taiwan}

\author{Alakesh Baishya}
\affiliation{Theoretical Quantum Physics Laboratory, RIKEN Cluster for Pioneering Research, Wako-shi, Saitama 351-0198, Japan}

\author{Yu-Ran Zhang}
\affiliation{Theoretical Quantum Physics Laboratory, RIKEN Cluster for Pioneering Research, Wako-shi, Saitama 351-0198, Japan}

\author{Anton Frisk Kockum}
\affiliation{Department of Microtechnology and Nanoscience, Chalmers University of Technology, 412 96 Gothenburg, Sweden}

\author{Yueh-Nan Chen}
\affiliation{Theoretical Quantum Physics Laboratory, RIKEN Cluster for Pioneering Research, Wako-shi, Saitama 351-0198, Japan}
\affiliation{Department of Physics and Center for Quantum Frontiers of Research \& Technology (QFort), National Cheng Kung University, Tainan 701, Taiwan}

\author{Fu-Li Li}
\email{flli@xjtu.edu.cn}
\affiliation{Ministry of Education Key Laboratory for Nonequilibrium Synthesis and Modulation of Condensed Matter, Shaanxi Province Key Laboratory of Quantum Information and Quantum Optoelectronic Devices, School of Physics, Xi'an Jiaotong University, Xi'an, Shannxi, 710049, China}

\author{Jaw-Shen Tsai}
\affiliation{Department of Physics, Tokyo University of Science, Shinjuku, Tokyo 162-0825, Japan}
\affiliation{RIKEN Center for Quantum Computing, Wako-shi, Saitama 351-0198, Japan}

\author{Franco Nori}
\email{fnori@riken.jp}
\affiliation{Theoretical Quantum Physics Laboratory, RIKEN Cluster for Pioneering Research, Wako-shi, Saitama 351-0198, Japan}
\affiliation{RIKEN Center for Quantum Computing, Wako-shi, Saitama 351-0198, Japan}
\affiliation{Department of Physics, The University of Michigan, Ann Arbor, 48109-1040 Michigan, USA}

\date{\today}

\begin{abstract}

One-way quantum computing is a promising candidate for fault-tolerant quantum computing. Here, we propose new protocols to realize a deterministic one-way CNOT gate and one-way $X$-rotations on current quantum-computing platforms. By applying a delayed-choice scheme, we overcome a limit of most currently available quantum computers, which are unable to implement further operations on measured qubits or operations conditioned on measurement results from other qubits. Moreover, we decrease the error rate of the one-way logic gates, compared to the original protocol using local operations and classical communication. In addition, we apply our deterministic one-way CNOT gate in the Deutsch-Jozsa algorithm to show the feasibility of our proposal. We demonstrate all these one-way gates and algorithms by running experiments on the cloud quantum-computing platform IBM Quantum Experience.

\end{abstract}

\maketitle



\section{Introduction}
\label{sec:Introduction}

Measurement-based one-way quantum computing~\cite{Raussendorf2001, Raussendorf2003, Briegel2009}, where single-qubit operations and measurements are applied to an initial highly entangled cluster state or graph state~\cite{Briegel2001, Ghne2009}, is considered a promising candidate for quantum computation. This approach to quantum computing is potentially more robust against noise and errors than the quantum circuit model, due to advantages in implementing fault-tolerant quantum computing~\cite{Zwerger2013, Zwerger2014, Harper2019,Hastrup2021}. Recently, several proposals for combining one-way quantum computing with quantum error correction and quantum algorithms have been presented~\cite{Raussendorf2006,Varnava2006,Herr2018,Ferguson2021}. Unlike circuit-based quantum computing, in which the scale of the quantum computer is limited by the number of available qubits and gates, measurement-based quantum computing is restricted by the scale of the cluster state one can generate~\cite{Ferguson2021}. Large two-dimensional cluster states have been generated in experiments with photonic qubits~\cite{Asavanant2019, Larsen2019,Vasconcelos2020,Ndagano2020,Hilaire2021resource}.

The ideas of cluster states and one-way quantum computing were proposed two decades ago. The earliest works on one-way quantum computing and cluster states applied to solid state qubits and superconducting circuits were considered in Refs.~\cite{Tanamoto2006, You2007, Tanamoto2009}. Moreover, a large variety of protocols on a series of one-way logic gates have been proposed and implemented, including single-qubit, two-qubit, and multi-qubit logic gates, e.g., the Toffoli gate and the CARRY gate~\cite{Kay2006, Joo2007, Miwa2009, Wang2010, Shen2012, Bell2013, Hao2014, Arriagada2018}. Moreover, there have been a number of experiments generating cluster states, and constructing one-way quantum logic gates, performed in optical systems~\cite{Walther2005, Kiesel2005,Prevedel2007, Tame2007, Chen2007, Vallone2008, Kaltenbaek2010, Bell2013, Tame2014, Asavanant2019, Larsen2019}, trapped ions~\cite{Lanyon2013}, and nuclear magnetic resonance (NMR) systems~\cite{Ju2010}.

Recently, superconducting qubits (e.g.~\cite{You2011, Buluta2011, Gu2017, Kockum2019a, Kjaergaard2020,Yost2020}) have attracted significant attention for quantum computing due to their scalability and controllability. The superconducting platform is considered as one of the most promising for building universal quantum computers, as exemplified by, e.g., commercial devices, including IBM's~\cite{IBMQ, Kandala2017,Kandala2019,Chiesa2019} and Google's~\cite{Arute2019, Neill2021} superconducting processors. A series of theoretical and experimental protocols for generating cluster states of superconducting qubits have been presented in Refs.~\cite{Tanamoto2006, Zhang2006, You2007, Chen2007_2, Tanamoto2009, Li2016, Lahteenmaki2016_2, Yang2017, Wang2018, Gong2019, Ku2020}, as well as theoretical protocols~\cite{Raussendorf2003, Arriagada2018} for one-way logic gates in superconducting processors. However, experimental demonstrations of one-way quantum computing in superconducting systems were lacking until very recently~\cite{Besee2020, Shirai2021}.

In this work, we demonstrate deterministic one-way quantum computing using the IBM Quantum Experience (IBMQ). The IBMQ contains a series of quantum chips composed of superconducting qubits, which can be used to construct quantum logical circuits consisting of quantum logic gates. As an on-cloud quantum computing service, IBMQ provides users with a powerful platform for experimentally implementing not only fundamental quantum phenomena~\cite{ Smith2019, Perez2020, White2020,Huang2021} but also quantum information processing tasks~\cite{Devitt2016, Kandala2017, Chiesa2019, Woerner2019, Tacchino2019, YeterAydeniz2020,Mooney2021}. However, until very recently~\cite{Corcoles2021}, due to the limitation of the physical structure of the IBMQ chips, they did not provide sufficient functionality to directly test many concepts of one-way quantum logic gates. For instance, the IBM devices were not able to enact qubit operations conditioned on the results of qubit measurements. To the best of our knowledge, this remains the case for other quantum-computing systems.

\begin{figure}
\begin{center}
\includegraphics[width=0.8\linewidth]{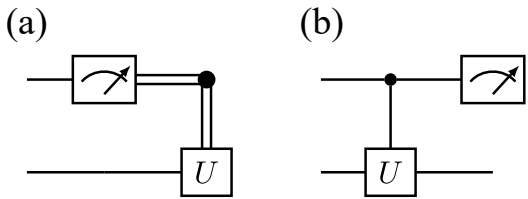}
\end{center}
\caption{Circuits for (a) a classically conditioned (the double line corresponds to classical communication) operation $U$ and (b) the delayed-choice approach for the same operation. We consider a situation where $U$ only acts on the second qubit when the outcome of the measurement on the first qubit is $1$. For such $U$, these two circuits yield the same outcome.
\label{delayedchoice}}
\end{figure}

We therefore introduce a protocol based on replacing the quantum operations conditioned on the measurement results [the ``local operations and classical communication'' (LOCC) approach] with delaying the choice of the measurement (the delayed-choice approach, see \figref{delayedchoice}). 
Although we need additional entangling gates to implement this delayed-choice approach, it opens up for experimental tests of many important concepts in one-way quantum computing.

We experimentally construct a deterministic one-way two-qubit gate (the one-way CNOT gate) and also use it to implement the Deutsch-Jozsa algorithm~\cite{Deutsch1985, Deutsch1992} on a 5-qubit chip called IBMQ Lima. Note that here ``deterministic'' means that the one-way logic gate has a fixed output; for a ``non-deterministic" one-way logic gate, a feedforward operation is required after measuring and obtaining the output. In the following, as we are not going to discuss non-deterministic one-way logic gates, for brevity we will omit ``deterministic''.

For the one-way CNOT gate, the measurement bases on the cluster state are fixed, while for most single-qubit one-way logic gates, the measurement bases of one subsystem are determined by the measurement results from another subsystem (LOCC). Therefore, implementing a one-way logic gate requires performing measurements at different times, and the feedback of measurement results can also introduce errors~\cite{Raussendorf2001,Raussendorf2003}. By applying the delayed-choice method, one-way logic gates can be achieved with simultaneous measurements on the entire cluster state. We can thus significantly decrease the errors from measurements and feedback operations. As an explicit example, we investigate the one-way $X$-rotation gate. With our approach, superconducting qubits become more amenable to one-way quantum computing.

This paper is organized as follows. In Sec.~\ref{sec:one-way CNOT gate}, we introduce the replacement of the quantum operation conditional on the classical measurement result with delayed choice of the measurement. With this approach, we can experimentally realize a one-way CNOT gate. We also apply the one-way CNOT gate in the Deutsch-Jozsa algorithm on the 5-qubit IBMQ Lima. In Sec.~\ref{sec:one-way rotation X}, we show the advantage of considering the delayed choice of the measurement and experimentally present the one-way $X$-rotation gate. Finally, in Sec.~\ref{sec:SummaryDiscussion}, we briefly summarize our results and discuss limitations for the possibility of developing one-way quantum computing on quantum processors. 


\section{Construction and application of the one-way CNOT gate}
\label{sec:one-way CNOT gate}


\subsection{The standard one-way CNOT gate}
\label{sec:standard one-way CNOT gate}

\begin{figure}
\begin{center}
\includegraphics[width=0.8\linewidth]{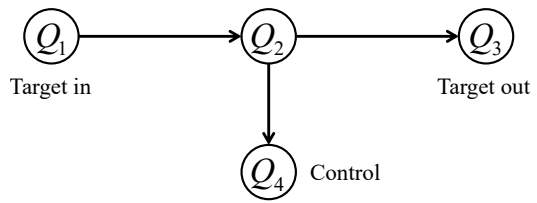}
\end{center}
\caption{Schematic illustration of the setup for the one-way CNOT gate. The full system consists of four qubits. Qubits $Q_1$, $Q_3$, and $Q_4$ act as the input target state, the output target state, and the control, respectively. The arrows represent standard CZ gates, which starts at the control qubit and ends at the target qubit.
\label{1wcnot}}
\end{figure}

We start by briefly recalling how to construct the cluster state for implementing a one-way CNOT gate~\cite{Raussendorf2001,Raussendorf2003}. The system, consisting of four qubits ($Q_1$, $Q_2$, $Q_3$, and $Q_4$; see \figref{1wcnot}), is first prepared in the state
\begin{equation}
\ket{\Psi}_{\rm in}^{\rm CNOT} = \ket{\psi}_1 \ket{+_x}_2 \ket{+_x}_3 \ket{\phi}_4,
\label{Eq:initial state}
\end{equation}
where $\ket{\pm_x}$ is the eigenstate of the Pauli $X$ matrix with the eigenvalue $\pm 1$. The subscripts in \eqref{Eq:initial state} are used to denote the physical qubits. Here, $\ket{\psi}_1$ and $\ket{\phi}_4$ are the input target state and control state, respectively, of the CNOT gate, while $Q_3$ plays the role of the output target state. More specifically, if $\ket{\phi}_4 = \ket{1}_4$, the $X$ operation is carried out on the ``transferred'' state $\ket{\psi}_3$.

\begin{figure*}
\begin{center}
\includegraphics[width=0.8\linewidth]{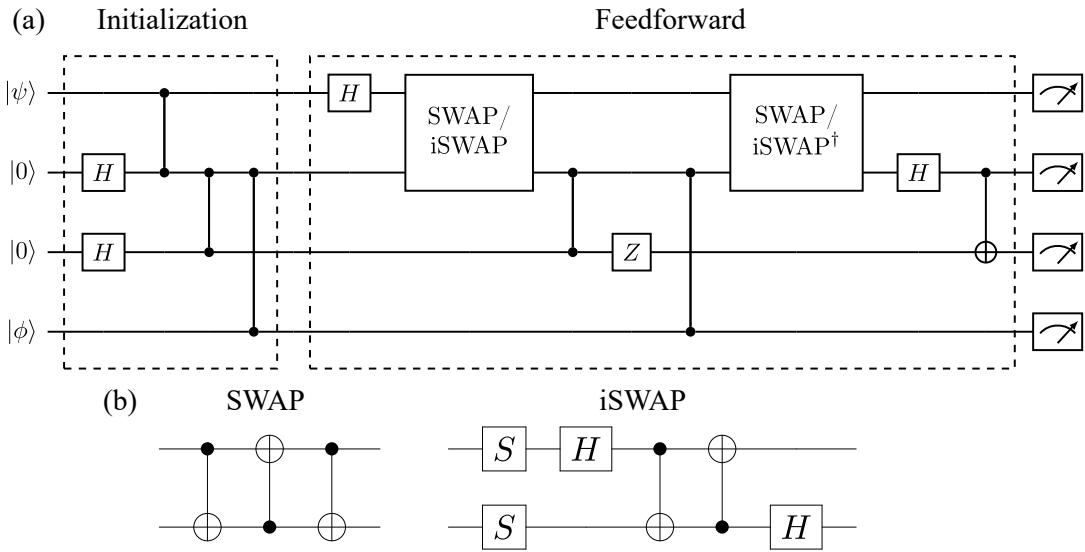}
\end{center}
\caption{(a) The circuit for constructing the delayed-choice one-way CNOT gate on the IBMQ Lima with the first and fourth qubits being the input target and the control, respectively. The output target is the third qubit. Here the SWAPs or iSWAPs are inserted due to the limited connectivity of the IBMQ Lima. (b) The gate decomposition of the SWAP gate and the iSWAP gate into CNOTs (the native two-qubit gate of IBMQ devices) and single-qubit gates.
\label{Limacnot}}
\end{figure*}

The above initialization process can be achieved by implementing suitable single-qubit logic gates. After the initialization, we sequentially apply a series of CZ gates $C_Z^{(12)}$, $C_Z^{(23)}$, and $C_Z^{(24)}$ to prepare the cluster state~\cite{Raussendorf2003,Arriagada2018}. Here, $C_Z^{(ij)}$ represents a CZ gate with $i$ and $j$ labelling the control and target qubits, respectively. The full initialization circuit is shown in the left part of \figpanel{Limacnot}{a}.

After constructing the cluster state, we simultaneously measure $X$ on $Q_1$ and $Q_2$ with the results $s_1$ and $s_2$, respectively. Here, we use the notation $s_i = 0$ ($s_i = 1$) to denote the corresponding outcome of the $i$th qubit with the eigenvalue $+1$ ($-1$) of the Pauli $X$ matrix. Finally, to realize deterministic one-way quantum computing, a feedforward operation,
\begin{equation}
U^{\rm CNOT}_\Sigma = Z_3^{s_1 + 1} X_3^{s_2} \otimes Z_4^{s_1},
\label{Eq:feedforward operation}
\end{equation}
is applied on $Q_3$ and $Q_4$. Here, $\alpha_j$ ($\alpha = X, Y, Z$) denotes the corresponding Pauli operation acting on qubit $j$. Each operation in \eqref{Eq:feedforward operation} is conditioned on the measurement results, which are transferred with classical communication. Note that for implementing the one-way CNOT gate, the measurement bases of the subsystems are always fixed.


\subsection{Construction of the delayed-choice one-way CNOT gate with IBMQ Lima}
\label{sec:refined one-way CNOT gate}

In this section, we experimentally implement the delayed-choice one-way CNOT gate using the IBMQ Lima, which is a 5-qubit chip having similar layout as \figref{1wcnot}, with another ancilla qubit $Q_5$ added below $Q_4$. The circuit for the one-way CNOT gate on IBMQ Lima is shown in \figref{Limacnot}.

Although the standard method of realizing a one-way gate can be experimentally implemented in many physical systems, including optical systems, trapped ions, and NMR~\cite{Walther2005, Prevedel2007, Tame2007, Chen2007, Vallone2008, Kaltenbaek2010, Ju2010, Bell2013, Lanyon2013, Tame2014}, superconducting systems like IBMQ could not (until very recently~\cite{Corcoles2021}) directly apply the same strategy to achieve a one-way CNOT gate, due to a major limitation: Conditional operations, based on classical transfer of readouts, are forbidden. These ``missing'' functions prohibit the feedforward operation. To overcome these limitations, we apply the quantum delayed-choice method to replace the classical conditional operations.

In general, the quantum delayed-choice approach enables us to create a superposition of eigenstates, corresponding to all the possible readouts, and the superposed state collapses to the required eigenstate after the measurement operations. In this way, classical conditional operations can be replaced by CZ or CNOT operations, which are available on IBMQ devices. 
By relaxing the requirement that all operations following the initial entangling step should be single-qubit gates and measurements, we can now test many concepts and applications in the current IBM devices. Moreover, we can also analyze the effect of the error mitigation on the one-way computation (see also the later example). We note that the delayed-choice approach does not make any additional assumption of functionality for one-way quantum computing on the superconducting platform, since all the required operations can be carried out before measuring any qubit, such that we can consider this as a part of the cluster-state generation. 

We now describe in detail the circuit model for the delayed-choice one-way CNOT operation. As mentioned above, the cluster state can be constructed by sequentially applying $C_Z^{(12)}$, $C_Z^{(23)}$, and $C_Z^{(24)}$ on the initial state given in \eqref{Eq:initial state}. Before the delayed-choice operation, we first perform the Hadamard gate on $Q_1$ to transfer the measurement basis from $Z$ to $X$. Then, a SWAP or iSWAP gate is applied to exchange the quantum states of $Q_1$ and $Q_2$, due to the architecture of IBM Lima.

Note that a SWAP gate can be decomposed into three CNOT gates, while an iSWAP gate is constructed by two CNOT gates and four single-qubit gates, as shown in \figpanel{Limacnot}{b}. In this way, we can decrease the gate error by replacing SWAP gates with iSWAP gates, because the CNOT gate in general introduces more error than single-qubit gates.

Next, we implement the delayed choice of $U_1=Z_3^{s_1 + 1}$ and $U_3=Z_4^{s_1}$, by applying $C_Z^{(23)}$ followed by $Z_3$, and $C_Z^{(24)}$, respectively. After the delayed choice of $U_1$ and $U_3$, the inverse of the SWAP or iSWAP operation is performed on $Q_1$ and $Q_2$. Finally, a Hadamard gate, following the delayed choice of $U_2=X_3^{s_2}$, is implemented.

To study whether iSWAP gates can reduce the noise from gate errors and the intrinsic environment interaction, we consider three different scenarios:

($i$) the delayed-choice one-way CNOT gate with SWAP gates;

($ii$) the delayed-choice one-way CNOT gate with iSWAP gates; and

($iii$) the delayed-choice one-way CNOT gate with iSWAP gates and error mitigation.

Here, we use the simplest error mitigation method provided by IBMQ~\cite{Kandala2019,Chen2019,Dogra2021,Maciejewski2020,Suchsland2021algorithmicerror}.
In short, before experimentally implementing the delayed-choice CNOT gate, we test the results of $Z$ measurements with the inputs being its eigenstates. The difference between ideal and tested results provides an error matrix which is used for error mitigation. 

\begin{figure}
\begin{center}
\includegraphics[width=\linewidth]{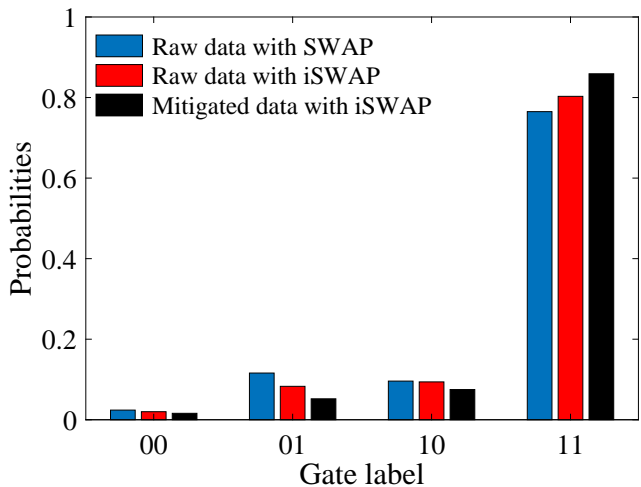}
\end{center}
\caption{Results of the one-way CNOT gate with inputs $\ket{0}$ (target) and $\ket{1}$ (control). The theoretical prediction of the output after the post-processing is $\ket{11}$. The blue, red, and black bars correspond to using SWAP, iSWAP, and iSWAP with error mitigation, respectively.
\label{cnotresult}}
\end{figure}

The experimental results from IBMQ Lima are presented in \figref{cnotresult}. To increase the transparency of the results, post-processing is performed by summing the results of $Q_1$ and $Q_2$, because ($i$) theoretically, the results of $Q_1$ and $Q_2$ are uniform in the delayed-choice one-way CNOT gate, and ($ii$) $Q_3$ and $Q_4$ play the roles of the output of the target qubit and the controlled qubit, respectively, in the standard CNOT gate. In other words, we only need to access the results of $Q_3$ and $Q_4$ to see whether the delayed-choice one-way CNOT gate functions successfully. 

Here, we only consider the case where the inputs of the control qubit and the target qubit are $\ket{1}$ and $\ket{0}$, respectively. Thus, the output of the target qubit should be $\ket{1}$. The experimental results for all the different inputs are shown in \appref{app:ExperimentalData}. We find that the experimental result with iSWAP gates outperforms the one with SWAP gates, which we expected since the sequence with iSWAP gates is less likely to introduce errors, as discussed above. Moreover, the error mitigation further improves the experimental results. In addition, we also apply quantum process tomography to compute the gate fidelity in terms of the Choi representation, namely
\begin{equation}
    F(G_1,G_2)=\text{tr}(\sqrt{\sqrt{\rho_{G_1}}\rho_{G_2}\sqrt{\rho_{G_1}}})^2,
\end{equation}
where
\begin{equation}
\rho_{G_i}=(\openone\otimes G_{i})\ket{\Psi}\bra{\Psi},
\end{equation}
is the Choi state of the quantum operation $G_i$ with $\ket{\Psi}=\sum_i 1/\sqrt{d}\ket{ii}$ being the maximally entangled state. Here, $d$ is the dimension of the local system. The experimental fidelities are, respectively, 0.804, 0.808, and 0.857, for the three different scenarios.

The device parameters of the IBMQ Lima superconducting processor used in our experiment are as follows. The average $T_1$ and $T_2$ of IBMQ Lima are  95.86~$\mu$s and 99.21~$\mu$s, which represents the energy decay time and dephasing time of the qubits, respectively. The average two-qubit gate error rate $\lambda$ (also called the two-qubit Clifford gate error rate) and gate time are approximately 1.1$\%$ and 405~ns, respectively. The average readout error is approximately 3$\%$ with a readout time around $\unit[5]{\mu s}$. In our simulation model, the error, introduced by the two-qubit gate, plays a significant role in our system. This is because the gate time and gate error of the two-qubit gate are both one order of magnitude greater than those of the single-qubit gates. We note that these parameters vary over time, and the details of the parameters used in this work are presented in Appendix~\ref{app:ExperimentalData}.


Here, we analyze each aforementioned error by some well-established models. We note that each model within our analysis is also used in the quantum assembly (QASM) simulator provided by the IBMQ library. The impact of qubit decoherence on the system obeys the quantum master equation:
\begin{equation}
\begin{aligned}
\frac{d\rho}{dt}=\sum_{i=1}^4\frac{\kappa_1^i}{2}\left[2\sigma_-^i\rho\sigma_+^i-\sigma_+^i\sigma_-^i\rho-\rho\sigma_+^i\sigma_-^i\right] \\
+\sum_{i=1}^4\frac{\kappa_\phi^i}{2}\left[2\sigma_z^i\rho\sigma_z^i-\sigma_z^{i^2}\rho-\rho\sigma_z^{i^2}\right],
\end{aligned}
\end{equation}
where $\rho$ is the density matrix, $\sigma_+^i$ ($\sigma_-^i$) denotes the creation (annihilation) operator, and $\sigma_z^i$ represents the Pauli $Z$ operator.
Here, the parameters $\kappa_1^i=1/T_1^i$ and $\kappa_\phi^i=1/T_\phi^i$ are determined by the relaxation and pure dephasing times, respectively, where $1/T_2 = 1/(2 T_1) + 1/T_\phi$. The delayed-choice one-way CNOT gates with SWAP gates and iSWAP gates require 12 and 10 two-qubit gates, respectively, and all the qubits are measured simultaneously. Thus, taking readout time into consideration, the total time to carry out the one-way CNOT gate is approximately 10~$\mu$s. This brings the system a decoherence error rate of around 4$\%$.

Here, we consider the effect of the gate error. In the quantum assembly simulator, the gate error of an $n$-qubit gate is described by an $n$-qubit depolarization error:
\begin{equation}
E_G(\lambda_G)\rho=(1-\lambda_G)\rho+\lambda_G\frac{I}{2^n},
\end{equation}
where $\lambda_G$ is the error rate. Inserting the number of CNOT gates and the two-qubit error rate, we obtain a error of approximately 10$\%$. We note that the gate error is deduced from randomized benchmarking~\cite{Magesan2011,Magesan2012}, where the two-qubit gate error rate represents the imperfection of both the single-qubit and CNOT gates.


Finally, we consider the effect of the readout error. The single-qubit readout error on a qubit can be described by probabilities $\lambda_{01}$ and $\lambda_{10}$. Here, $\lambda_{01}$ ($\lambda_{10}$) is the probability of recording a noisy measurement outcome as $1$ ($0$) given the ideal measurement outcome was $0$ ($1$). 
According to the simulation result on the QASM simulator, the readout error of the two output qubits leads to an additional error rate of approximately 10$\%$. We note that the above three models are considered independently by inserting the system's parameters into the simulations.

As a result, the main factors of noises, interfering with the operation of the delayed-choice one-way CNOT gate, are gate error and readout error. Either one leads to an error rate of around 10$\%$. We recall that the decoherence of qubits causes an additional error rate of around 4$\%$. Moreover, the total error can be reduced by applying measurement error mitigation. Our simulation fits the corresponding experimental results in Table.~\ref{tab:cnot}. All the estimated error rates are obtained via simulation on the QASM simulator, using the parameters of the IBMQ Lima.





\begin{figure}
\begin{center}
\includegraphics[width=\linewidth]{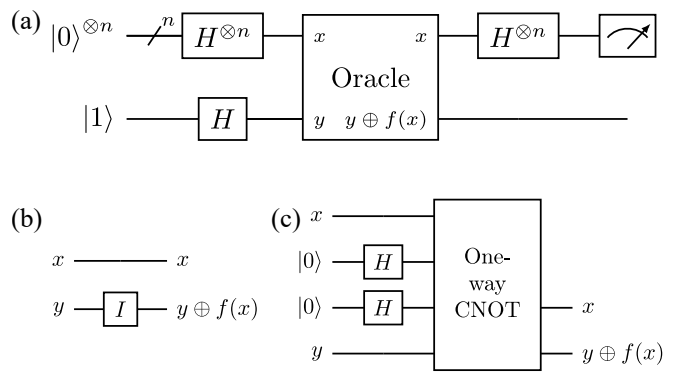}
\end{center}
\caption{Circuits for the Deutsch-Jozsa (DJ) algorithm with the delayed-choice one-way CNOT gate.
(a) The fundamental circuit representation of the $n$-bit Deutsch-Jozsa algorithm. The first $n$ qubits are all prepared in $\ket{0}$ while the ancilla qubit is prepared in $\ket{1}$.
(b) The circuit representation of a $1$-bit Boolean function with $f(x) = 0$.
(c) The circuit representation of a $1$-bit Boolean function with $f(x) = x$.
\label{DJA}}
\end{figure}

\subsection{Implementing the Deutsch-Jozsa algorithm using the delayed-choice one-way CNOT gate}

After successfully constructing the delayed-choice one-way CNOT gate, we implement a practical Deutsch-Jozsa (DJ) quantum algorithm~\cite{Deutsch1985, Deutsch1992} with this operation to show how our one-way CNOT gate works in practical problems.
%

The Deutsch-Jozsa algorithm is the first quantum algorithm ever invented. It finds out whether an $n$-bit Boolean function $f(x): \mleft\{ 0, 1 \mright\}^n \rightarrow \mleft\{ 0, 1 \mright\}$ with $\{ x \in R | 0 \leq x \leq 2^{n - 1} | n \}$ is balanced or constant. Here, a constant function means that the result remains unchanged with arbitrary input, while for a balanced function half of the possible inputs yield the result $0$, and the other half yield the result $1$. A classical algorithm requires $(2^{n - 1} + 1)$ queries to distinguish the two types of functions, but the DJ algorithm can complete the process with a single query.

The logical process of the DJ algorithm is as follows [see also \figpanel{DJA}{a}]:
To classify an $n$-bit Boolean function, we first initialize all $n$ qubits (representing the input $x$ of the function) in $\ket{0}$. An ancilla qubit $q_{n+1}$ is prepared in $\ket{1}$. Then, Hadamard gates are applied on all qubits. Denoting the state of the ancilla qubit by $\ket{y}$, an oracle is implemented which maps $\ket{x} \ket{y}$ to $\ket{x} \ket{y \oplus f(x)}$. Finally, we measure all $n$ qubits in the $X$ basis, which is achieved by inserting Hadamard gates before measuring $Z$. If the final outcome sums up to $1$, the corresponding function $f(x)$ is balanced, otherwise the function is constant. The explicit circuits for the constant function $f(x) = 0$ and the balanced function $f(x) = x$, using the delayed-choice one-way CNOT gates, are presented in \figpanel{DJA}{b} and \figpanelNoPrefix{DJA}{c}, respectively.

\begin{figure}
\begin{center}
\includegraphics[width=0.8\linewidth]{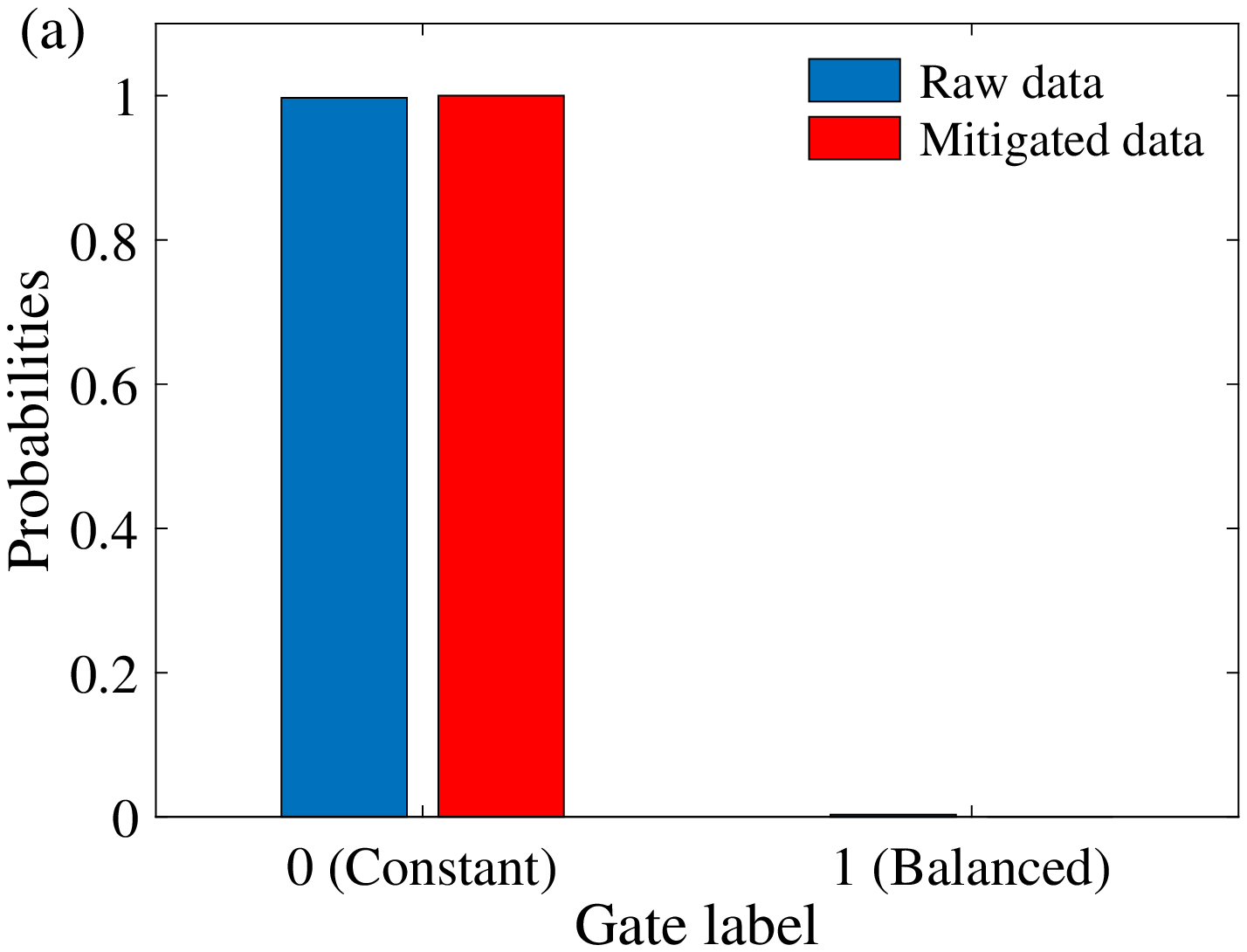}
\includegraphics[width=0.8\linewidth]{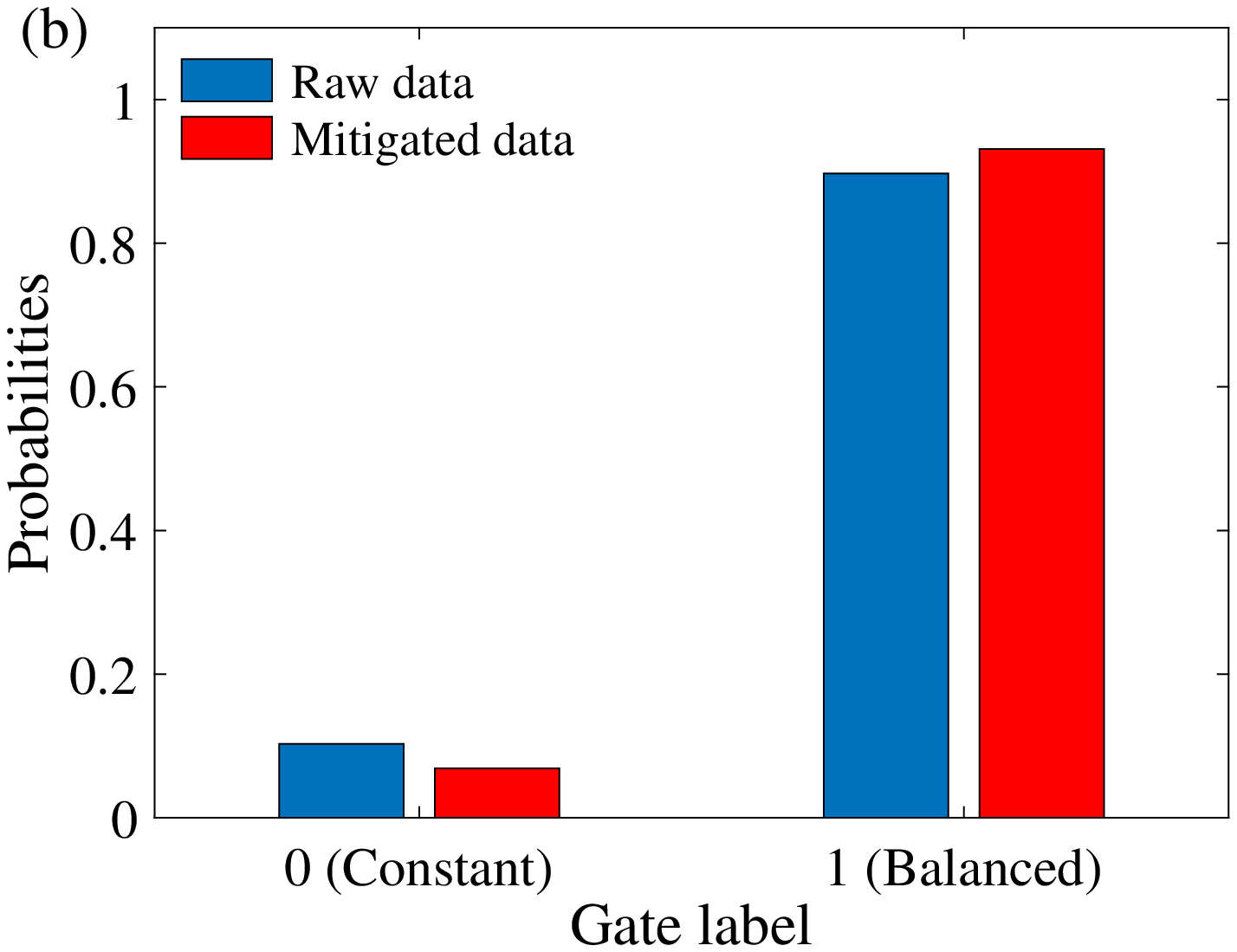}
\end{center}
\caption{Results for running the Deutsch-Jozsa algorithm on IBMQ Lima with (a) the constant function $f(x) = 0$ and (b) the balanced function $f(x) = x$. In (b), we apply the delayed-choice one-way CNOT gate in the oracle [see \figpanel{DJA}{c}]. The blue and red bars correspond to results with and without measurement-error mitigation, respectively. When the measurement result is $0$, the function is classified as constant. Otherwise it is classified as balanced.
\label{DJAresult}}
\end{figure}

Figure~\ref{DJAresult} shows experimental results for distinguishing the balanced [$f(x) = 0$] and constant [$f(x) = x$] functions, respectively, by the DJ algorithm with the delayed-choice one-way CNOT gate. These results are also obtained on IBMQ Lima. Although the probability distributions are not equivalent to the theoretical predictions due to inevitable noise, our experimental results still show how the DJ algorithm works for the discrimination problem. In particular, for the balanced function, while the success probability $p(0) < 1$, it is significantly higher than the probability $p(1)$ of the wrong answer. We note that, for the DJ algorithm, one would not need the success probability to be unity. In other words, the balanced and constant functions can be discriminated as long as the corresponding success probability is higher than the fail probability. Our experimental results satisfy this criterion by a wide margin.


\section{Construction of the delayed-choice one-way $X$-rotation gate}
\label{sec:one-way rotation X}

As we attempt to develop a universal set of one-way logic gates, we need to construct single-qubit rotation gates, in addition to the entangling two-qubit gate demonstrated in Sec.~\ref{sec:one-way CNOT gate}. For such single-qubit gates, the measurement bases of one subsystem are determined by the measurement results of another subsystem. Therefore, implementing a standard one-way logic gate, in general, requires sequential measurements and the feedback of measurement results. In addition, due to the non-instantaneous nature of the measurements, which usually are slower than a CZ gate, the feedback of the measurement results also introduces more errors. In this part, we consider the one-way $X$-rotation gate as a concrete example, to demonstrate our method to construct one-way rotation gates with delayed measurements.


\begin{figure}
\begin{center}
\includegraphics[width=0.8\linewidth]{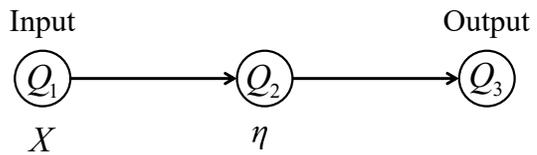}
\end{center}
\caption{The basic idea of the one-way $X$-rotation gate. The qubits $Q_1$ and $Q_3$ are the input and output qubits, respectively, and the arrows represent the standard CZ gates, which start at the control qubit and end at the target qubit. Here, the $X$ under qubit $Q_1$ represents that qubit $Q_1$ is measured in the $X$ basis, and the $\eta$ under qubit $Q_2$ represents the measurement basis of qubit $Q_2$, which is in the $xy$ plane with an angle $\eta$ to the $x$ axis.}
\label{xrot}
\end{figure}

Before introducing the delayed-choice one-way $X$-rotation gate, we briefly recall how to achieve the standard one-way $X$-rotation gate
\begin{equation}
\label{Eq:xrot}
U_{x}(\alpha) = 
\begin{pmatrix}
\cos(\frac{\alpha}{2}) & -i\sin(\frac{\alpha}{2}) \\
-i\sin(\frac{\alpha}{2}) & \cos(\frac{\alpha}{2})
\end{pmatrix}
,
\end{equation}
with a rotation angle $\alpha$. We first initialize the three-qubit system ($Q_1$, $Q_2$, and $Q_3$), shown in \figref{xrot}, in the state
\begin{equation}
\ket{\Phi}_{\rm in}^x = \ket{\phi}_1 \ket{+_x}_2 \ket{+_x}_3,
\end{equation}
where $\ket{\phi}_1$ is the input state of the gate. After the initialization, a cluster state is constructed by applying the CZ gates $C_Z^{(12)}$ and $C_Z^{(23)}$ on the system. Then, an $X$ measurement is performed on $Q_1$ with outcome $s_1$. Now, the measurement basis for $Q_2$ is determined by $\eta = (-1)^{s_1 + 1} \alpha$. Here, $\eta$ is the phase angle on the $xy$ plane of the measurement basis. As we can see, the feedback $s_1$ will decide the measurement basis. Finally, a feedforward operation
\begin{equation}
U^X_\Sigma = Z_3^{s_1} X_3^{s_2},
\label{Eq:ff_xrot}
\end{equation}
is applied to $Q_3$, changing the state of $Q_3$ to $U_{x}(\alpha)\ket{\phi}_3$.

\begin{figure*}
\begin{center}
\includegraphics[width=0.8\linewidth]{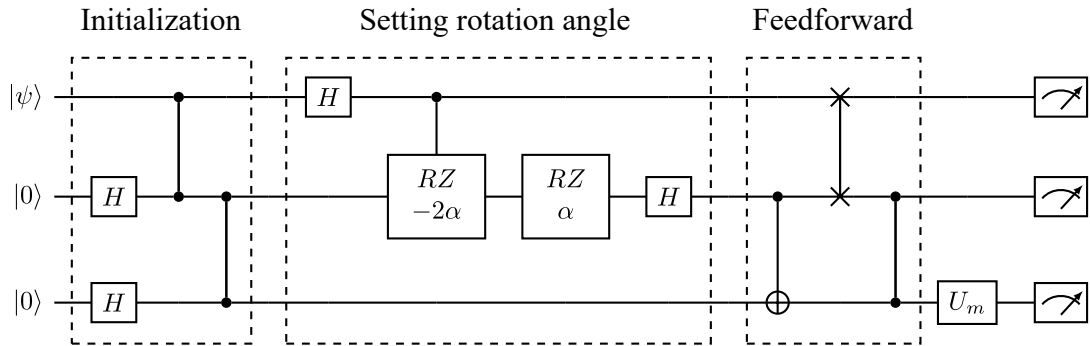}
\end{center}
\caption{The circuit for the delayed-choice one-way $X$-rotation gate. The input and output states are encoded in qubits $Q_1$ and $Q_3$, respectively. The gate $U_m$ before measuring qubit $Q_3$ represents a rotation operation that decides the measurement basis for $Q_3$.}
\label{xrotcirc}
\end{figure*}

In what follows, we apply the delayed-choice method to the one-way $X$-rotation gate. The quantum circuit model for the delayed-choice one-way $X$-rotation gate is shown in \figref{xrotcirc}. In the feedforward part, we apply a controlled $Z$-rotation gate on $Q_1$ and $Q_2$ (denoted by $CRZ_{12}$ in the following) with an angle $- 2 \alpha$. The subscript $12$ corresponds to the control qubit $Q_1$ and the target qubit $Q_2$, respectively. After the $CRZ_{12}$ gate, we apply a single-qubit $Z$-rotation gate $RZ$ on $Q_2$ with an angle $\alpha$. These two gates generate a superposition state of two eigenstates of the projectors on the $xy$ plane with the angle $\eta = \pm \alpha$. Therefore, the superposed state contains the readout information $s_1 = 0$ and $s_1 = 1$, corresponding to different total rotation angles $\eta = \alpha$ and $\eta = - \alpha$.

After measuring $Q_1$, the superposed state collapses to the state with the corresponding result $s_1$. On the other hand, we also apply the delayed choice to carry out the feedforward operation in \eqref{Eq:ff_xrot}. Note that we do not need to replace the swapped qubit, if it is not involved in any further multi-qubit operation. In this protocol, one SWAP operation between $Q_1$ and Q$_2$ is sufficient, so we do not replace it with iSWAP gates.

\begin{figure}
\begin{center}
\includegraphics[width=0.85\linewidth]{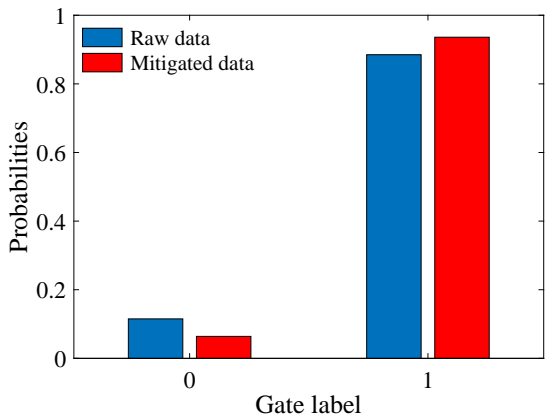}
\end{center}
\caption{Experimental results for the delayed-choice one-way $X$-rotation gate on IBMQ Lima. The blue and red bars correspond to conditions without and with measurement-error mitigation, respectively. We use the input state $\ket{0}_1$ and the rotation angle $\alpha = \pi/2$. Thus, the output should ideally be $\ket{-_y}_3$, and this is what the measurement result $1$ will correspond to.
\label{xrotresult}}
\end{figure}

Figure~\ref{xrotresult} shows the results of running the delayed-choice one-way $X$-rotation gate on IBMQ Lima with and without measurement-error mitigation. We consider the input on $Q_1$ to be $\ket{0}_1$ and the rotation angle is $\alpha = \pi / 2$. Therefore, the output state is expected to be rotated to $\ket{-_y}_3$, where $\ket{-_y}$ is the eigenstate of the Pauli $Y$ matrix with the eigenvalue $-1$. We also present the experimental results for other input states in \appref{app:ExperimentalData}. As shown in \figref{xrotresult}, the probability of obtaining the expected readout without error mitigation is $0.916$, while the mitigated result is $0.952$. In addition, we have performed process tomography of the one-way $X$-rotation gate and obtained a gate fidelity of $0.964$ ($0.948$) with (without) error mitigation.

The delayed-choice $X$-rotation gate requires eight CNOT gates and three physical qubits. Using the aforementioned error analysis in Sec.~\ref{sec:refined one-way CNOT gate}, the error rate introduced by pure dephasing and energy dissipation is approximately 1.5$\%$. Moreover, the eight CNOT gates for the one-way $X$-rotation gate combined together have a total error rate of approximately 3.5$\%$. Finally, readout error would lead to an additional error rate of approximately 2.6$\%$. Therefore, these three factors of error together would be the main limitations of our delayed-choice one-way $X$-rotation gate in these experiments. 

Compared to the delayed-choice one-way CNOT gate discussed in Sec.~\ref{sec:refined one-way CNOT gate}, the influence of all the three noise channels on the $X$-rotation gate decreases. This is because we avoided using a qubit with poor parameters in the $X$-rotation gate case.
In the one-way CNOT gate, the relaxation time $T_1$ and decoherence time $T_2$ of qubit $Q_3$ are around $2/3$, which is less than for the qubits $Q_0$-$Q_2$, and the readout error of $Q_3$ is approximately twice as much as the readout error of $Q_0$-$Q_2$. The gate error of two-qubit gates involving $Q_3$ is also twice the two-qubit gate errors between $Q_0$-$Q_1$ and $Q_1$-$Q_2$. Since we did not use $Q_3$ in the delayed-choice one-way $X$-rotation gate, the total error rates of all the three noise channels decreased significantly. Our simulation model also supports the above analysis.



\section{Summary and discussion}
\label{sec:SummaryDiscussion}

In this article, we have proposed and demonstrated a delayed-choice method to replace the classical conditional operations for realizing deterministic one-way logic gates. 
Our method can, in general, significantly decrease the impact of readout error and thus increase the fidelity of  one-way logic gates. With this simple approach, we can experimentally implement one-way quantum logical gates in noisy intermediate-scale quantum computers, e.g., the IBM Quantum Experience. We presented two explicit examples: ($i$) a one-way CNOT gate and ($ii$) a one-way $X$-rotation gate.

In our experiments on the IBMQ Lima, we showed how these delayed-choice one-way logic gates perform on currently available devices. We also applied the one-way CNOT gate in the Deutsch-Jozsa algorithm to demonstrate that our approach can be used in some practical problems. Moreover, we further decreased the gate error by replacing the SWAP gates with iSWAP gates, and decreased measurement errors by applying measurement-error mitigation.

Comparing to standard one-way logic gates, which only require a series of single-qubit measurements after preparing the cluster states, our protocols additionally imported a series of ancilla single-qubit and two-qubit logic gates, which makes the protocol more complex. However, as we mentioned above, the delayed-choice approach generates a superposition of eigenstates and collapses to the required eigenstate after the measurement operations. With this simple relaxation, we can demonstrate basic one-way quantum computing in superconducting circuits. Moreover, we note that our delayed-choice proposal can significantly decrease the effect of measurement error. In the original feedforward one-way quantum gates, readouts of all the qubits are required to resolve the feedforward operations. Measurement error of any qubit will result in an incorrect quantum operation and thus bring unreliable feedforward operations. However, our delayed-choice proposal only performs the measurements on the ancilla qubits after which the state of the system collapses and the correctness of the measurement results is not critical. Thus, in our delayed-choice proposal, the measurement errors on the ancilla qubits do not affect the fidelity of the one-way logical gate.

As we have discussed in Sec.~\ref{sec:one-way CNOT gate} and Sec.~\ref{sec:one-way rotation X}, the performance of delayed-choice one-way gates are mainly affected by qubit decoherence, gate error, and measurement error. In addition, one-way quantum computing is also limited by the scale of the cluster state we can construct, and currently the largest available superconducting quantum device contains only several dozens of qubits~\cite{Arute2019, Wu2021, IBMQ}. For practical applications, it is desirable to have more qubits, and future quantum devices are planned with many more qubits.

Finally, it would be interesting to complete universal one-way quantum gates, especially the remaining one-Hadamard and one-way single-qubit arbitrary rotation gates. However, due to the limitation of IBMQ Lima, many SWAP gates need to be inserted into the circuit models, such that the circuit depth increases drastically. We have presented details of the circuit models and the simulation results in Appendices~\ref{Hadamard} and \ref{arbitrary}.

\begin{acknowledgments}

We acknowledge the IBM Quantum Experience for providing us a platform to implement the experiment. The views expressed are those of the authors and do not reflect the official policy or position of IBM or the IBM Quantum Experience team. We also acknowledge fruitful discussions with R.~Wang, Y.~Zhou, D.-K.~Zhang, S.~Devitt, Z.-Q.~Yin, S.-L.~Ma, Y.~Xu, H.~Mukai, and S.~Shirai.

Z.-P.Y.~acknowledges the support of China Scholarship Council.
H.-Y.K.~acknowledges the support of the Ministry of Science and Technology, Taiwan (Grant No.~MOST 110-2811-M-006-546). 
Y.-R.Z. acknowledges the Japan Society for the Promotion of Science (JSPS) (via the Postdoctoral Fellowship Grant No.~P19326 and the KAKENHI Grant No.~JP19F19326).
A.F.K.~acknowledges support from the Japan Society for the Promotion of Science (BRIDGE Fellowship BR190501), the Swedish Research Council (grant number 2019-03696), and the Knut and Alice Wallenberg Foundation through the Wallenberg Centre for Quantum Technology (WACQT).
Y.-N.C. is supported partially by the National Center for Theoretical Sciences and Ministry of Science and Technology, Taiwan, Grants No. MOST 110-2123-M-006-001, and the Army Research Office (under Grant No. W911NF-19-1-0081).
J.T.~acknowledges support from Japan Science and Technology Agency (JST) (via the Moonshot R\&D Grant No.~JPMJMS2067 and the CREST Grant No.~JPMJCR1676), and the New Energy and Industrial Technology Development Organization (NEDO), Japan (Grant No.~JPNP16007).
F.N. is supported in part by: Nippon Telegraph and Telephone Corporation (NTT) Research, the Japan Science and Technology Agency (JST) [via the Quantum Leap Flagship Program (Q-LEAP), the Moonshot R\&D Grant Number JPMJMS2061, and the Centers of Research Excellence in Science and Technology (CREST) Grant No. JPMJCR1676], the Japan Society for the Promotion of Science (JSPS) [via the Grants-in-Aid for Scientific Research (KAKENHI) Grant No. JP20H00134, the Army Research Office (ARO) (Grant No. W911NF-18-1-0358), the Asian Office of Aerospace Research and Development (AOARD) (via Grant No. FA2386-20-1-4069), and the Foundational Questions Institute Fund (FQXi) via Grant No. FQXi-IAF19-06.

\end{acknowledgments}


\appendix


\section{Detailed data obtained from IBMQ for the delayed-choice one-way CNOT and $X$-rotation gates}
\label{app:ExperimentalData}

In this appendix, we first present the detailed data for the one-way logic gates that we ran on IBMQ Lima. The experimental data of the delayed-choice one-way CNOT gates and the delayed-choice one-way $X$-rotation gate are presented in Tables~\ref{tab:cnot} and \ref{tab:xrot}, respectively. The parameters which we used in our simulation model are presented in Table~\ref{tab:LimaCalibration}.

\begin{table}
\caption{Probability distribution for the delayed-choice one-way CNOT gate implemented on IBMQ Lima. Here, ``Input $\ket{ij}$'' represents that the input of the target state and the control state are $\ket{i}$ and $\ket{j}$, respectively. The $\ket{kl}$ in the second to fifth column denotes the measurement with results $kl$.
\label{tab:cnot}}

\begin{ruledtabular}
\begin{tabular}{lcccc}
Input $\ket{00}$ & $\ket{00}$ & $\ket{01}$ & $\ket{10}$ & $\ket{11}$ \\
\colrule
Raw/SWAP & 0.787 & 0.096 & 0.090 & 0.026 \\
Raw/iSWAP & 0.848 & 0.071 & 0.067 & 0.014 \\
Mitigated/iSWAP & 0.866 & 0.063 & 0.057 & 0.013
\end{tabular}
\end{ruledtabular}

\qquad

\begin{ruledtabular}
\begin{tabular}{lcccc}
Input $\ket{01}$ & $\ket{00}$ & $\ket{01}$ & $\ket{10}$ & $\ket{11}$ \\
\colrule
Raw/SWAP & 0.025 & 0.108 & 0.110 & 0.757 \\
Raw/iSWAP & 0.018 & 0.089 & 0.099 & 0.794 \\
Mitigated/iSWAP & 0.012 & 0.052 & 0.075 & 0.860
\end{tabular}
\end{ruledtabular}

\qquad

\begin{ruledtabular}
\begin{tabular}{lcccc}
Input $\ket{10}$ & $\ket{00}$ & $\ket{01}$ & $\ket{10}$ & $\ket{11}$ \\
\colrule
Raw/SWAP & 0.119 & 0.779 & 0.024 & 0.078 \\
Raw/iSWAP & 0.095 & 0.837 & 0.015 & 0.053 \\
Mitigated/iSWAP & 0.069 & 0.871 & 0.013 & 0.048
\end{tabular}
\end{ruledtabular}

\qquad

\begin{ruledtabular}
\begin{tabular}{lcccc}
Input $\ket{11}$ & $\ket{00}$ & $\ket{01}$ & $\ket{10}$ & $\ket{11}$ \\
\colrule
Raw/SWAP & 0.125 & 0.023 & 0.748 & 0.104 \\
Raw/iSWAP & 0.096 & 0.016 & 0.811 & 0.077 \\
Mitigated/iSWAP & 0.061 & 0.013 & 0.854 & 0.072
\end{tabular}
\end{ruledtabular}
\end{table}

\begin{table}
\caption{Noise parameters of the IBMQ Lima. Qubits $Q_0$, $Q_1$, $Q_2$, and $Q_3$ are utilized in the delayed-choice one-way CNOT gate, while $Q_0$, $Q_1$, and $Q_2$ are involved in delayed-choice one-way $X$-rotation gate.
\label{tab:lima}}
\begin{ruledtabular}
\begin{tabular}{lccccc}
Qubits & $T_1$($\mu s$) & $T_2$($\mu s$) & $\lambda_{01}$ & $\lambda_{10}$ & Pauli-$X$ error ($\times 10^{-4}$) \\
\colrule
$Q_0$ & 152.60 & 192.19 & 0.0252 & 0.0094 & 1.96 \\
$Q_1$ & 114.79 & 106.56 & 0.0230 & 0.0034 & 2.20 \\
$Q_2$ & 108.53 & 109.94 & 0.0304 & 0.0060 & 2.18 \\
$Q_3$ & 85.16 & 69.66 & 0.0474 & 0.0168 & 2.79 \\
$Q_4$ & 18.24 & 17.71 & 0.1112 & 0.0176 & 7.16
\end{tabular}
\end{ruledtabular}
\qquad
\begin{ruledtabular}
\begin{tabular}{lcc}
CNOT Control-Target & Error rate ($\times 10^{-2}$) & Gate time (ns) \\
\colrule
0-1 & 0.5629 & 305.8 \\
1-0 & 0.5629 & 341.3 \\
1-2 & 0.5937 & 334.2 \\
2-1 & 0.5937 & 298.7 \\
1-3 & 1.683 & 497.8 \\
3-1 & 1.683 & 462.2 \\
3-4 & 1.623 & 519.1 \\
4-3 & 1.623 & 483.6
\end{tabular}
\end{ruledtabular}\label{tab:LimaCalibration}
\end{table}

\begin{table}
\caption{Probability distribution for the delayed-choice one-way $X$-rotation gate on IBMQ Lima. Here, $\ket{\pm_y}$ denotes the $Y$ measurement with the corresponding outcome $\pm 1$.
\label{tab:xrot}}

\begin{ruledtabular}
\begin{tabular}{lccc}
Input $\ket{0}$ & $\ket{+_y}$ & $\ket{-_y}$ \\
\colrule
Raw & 0.084 & 0.916 \\
Mitigated & 0.048 & 0.952
\end{tabular}
\end{ruledtabular}

\qquad

\begin{ruledtabular}
\begin{tabular}{lccc}
Input $\ket{1}$ & $\ket{+_y}$ & $\ket{-_y}$ \\
\colrule
Raw & 0.926 & 0.074 \\
Mitigated & 0.932 & 0.068
\end{tabular}
\end{ruledtabular}
\end{table}

\section{Delayed-choice protocol for the one-way Hadamard gate}
\label{Hadamard}

In this section, we introduce a protocol to construct a one-way Hadamard gate by utilizing the delayed-choice approach. We also show the simulation results we obtained on the quantum assembly simulator for this protocol.

\begin{figure}
\begin{center}
\includegraphics[width=\linewidth]{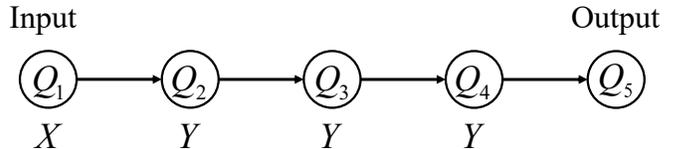}
\end{center}
\caption{The basic idea of the one-way Hadamard gate. The qubits $Q_1$ and $Q_5$ are the input and output qubits, respectively. The $X$ or $Y$ under each qubit represents the measurement basis used for those qubits.}
\label{Hbrief}
\end{figure}

The original idea for the one-way Hadamard gate is shown in Fig.~\ref{Hbrief}. The one-way Hadamard gate requires 5 physical qubits, labelled $Q_1$ to $Q_5$. $Q_1$ serves as the input qubit, while the other four qubits are initialized in the state $\ket{+_x}$. After constructing a linear cluster state on these 5 qubits by carrying out CZ gates $C_Z^{(12)}$, $C_Z^{(23)}$, $C_Z^{(34)}$, and $C_Z^{(45)}$, qubit $Q_1$ is measured in the $X$ basis, while qubits $Q_2$, $Q_3$, and $Q_4$ are measured in the $Y$ basis. These measurement operations can be carried out simultaneously, and we denote the measurement results $s_1$ to $s_4$, respectively. Finally, a feedforward operation $U_\Sigma^H=X^{s_1+s_3+s_4}Z^{s_2+s_3}$ is applied to the output qubit $Q_5$, and the output state corresponding to the Hadamard gate having acted on the input state is obtained.

\begin{figure*}
\begin{center}
\includegraphics[width=0.8\linewidth]{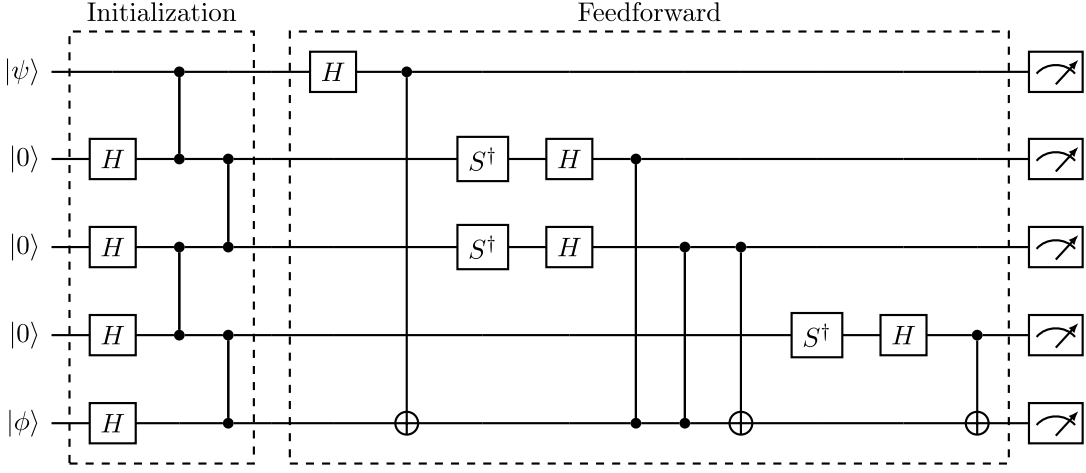}
\end{center}
\caption{The circuit for the delayed-choice one-way Hadamard gate. The input and output states are encoded in qubits $Q_1$ and $Q_5$, respectively. This circuit did not consider physical connectivity between qubits in actual hardware.}
\label{hadamardcirc}
\end{figure*}

Similar to the one-way CNOT gate, we can also apply the delayed-choice approach to carry out a one-way Hadamard gate on the IBMQ devices. The logical circuit of our one-way Hadamard gate is shown in \figref{hadamardcirc}. We note that this circuit does not take system connectivity into consideration. In the real device, we need to apply SWAP or iSWAP gates in order to demonstrate the quantum operations between non-neighboring qubits.



The simulation results for our one-way Hadamard gate, obtained on the quantum assembly simulator, are shown in Table~\ref{tab:Hadamard}. We not only consider the ideal case in which there is no noise in the circuit, but also perform simulations including noise corresponding to single-qubit and two-qubit gate fidelities of \unit[99.8]{\%} and \unit[98]{\%}, respectively. These gate errors are higher than those of IBMQ Lima. The results for the noisy case do not show very high fidelity, but are clearly enough to correctly identify the logical output. 

We also attempted to implement our circuit directly on IBMQ Lima, but the experimental results were clearly worse than the simulations had indicated and are not shown here. In order to realize a delayed-choice one-way Hadamard gate in IBMQ Lima, we need to apply at least six SWAP gates. In other words, it requires at least 27 CNOT gates, which is over twice that in the delayed-choice one-way CNOT gate. Such a large circuit depth gives decoherence too much time to destroy the quantum state.

\begin{table}
\caption{Probability distribution for the delayed-choice one-way Hadamard gate on the quantum assembly simulator. This circuit is designed for IBMQ Lima. This table shows the simulation results under two conditions:
(a) Ideal conditions without noise;
(b) Conditions with noise, using SWAP gates in the circuit in \figref{hadamardcirc} to compensate for the limited qubit connectivity.
For the condition (b), the fidelities of the single-qubit and two-qubit gates are set to \unit[99.8]{\%} and \unit[98]{\%}, respectively.
\label{tab:Hadamard}}

\begin{ruledtabular}
\begin{tabular}{lccc}
Input $\ket{+_x}$ & $\ket{0}$ & $\ket{1}$ \\
\colrule
Ideal & 1 & 0 \\
Noisy & 0.825 & 0.175
\end{tabular}
\end{ruledtabular}

\qquad

\begin{ruledtabular}
\begin{tabular}{lccc}
Input $\ket{-_x}$ & $\ket{0}$ & $\ket{1}$ \\
\colrule
Ideal & 0 & 1 \\
Noisy & 0.185 & 0.815
\end{tabular}
\end{ruledtabular}
\end{table}


\section{Delayed-choice protocol for one-way arbitrary single-qubit rotations}
\label{arbitrary}

The delayed-choice approach can also be utilized to construct a one-way arbitrary rotation gate, achieving a universal one-way logic gate set together with the one-way CNOT gate shown in Sec.~\ref{sec:one-way CNOT gate}. However, due to the limitations of the IBMQ devices, we were unable to obtain satisfying results on any real device. In this section, we therefore demonstrate the delayed-choice one-way arbitrary rotation gate with the quantum assembly simulator.

\begin{figure}
\begin{center}
\includegraphics[width=\linewidth]{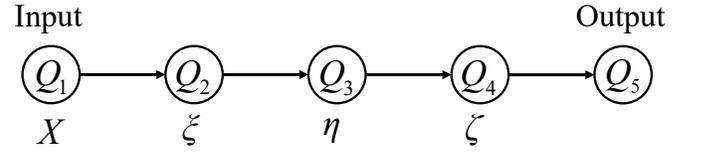}
\end{center}
\caption{The basic idea of the one-way arbitrary rotation gate. The qubits $Q_1$ and $Q_5$ are the input and output qubits, respectively. The $X$, $\xi$, $\eta$ and $\zeta$ under qubits $Q_1$ to $Q_4$ represent the measurement directions of the corresponding qubits.}
\label{Rbrief}
\end{figure}

The basic principle of a one-way arbitrary rotation gate $U_{\rm rot}(\xi,\eta,\zeta) = U_x(\zeta) U_z(\eta) U_x(\xi)$ is shown in Fig.~\ref{Rbrief}, where $U_a(\theta)$ represents a rotation operation about the $a$ axis ($a = x, z$) with rotation angle $\theta$. This gate is implemented using 5 physical qubits, labelled $Q_1$ to $Q_5$. Similar to the one-way $X$-rotation gate, $Q_1$ is the input qubit, while $Q_2$ to $Q_5$ are initialized in the state $\ket{+_x}$. After entangling the qubits with CZ gates $C_Z^{(12)}$, $C_Z^{(23)}$, $C_Z^{(34)}$, and $C_Z^{(45)}$, qubit $Q_1$ is measured in the $X$ basis, yielding the result $s_1$. After this, measurements on qubits $Q_2$, $Q_3$, and $Q_4$ are carried out in sequence, giving the corresponding results $s_2$, $s_3$, and $s_4$. Here, the measurement bases are in the $xy$ plane, with the angles to the $x$ axis $\alpha=(-1)^{s_1}\xi$, $\beta=(-1)^{s_2}\eta$, and $\gamma=(-1)^{s_1+s_3}\zeta$. Finally, a feedforward operation $U_\Sigma^{rot}=Z_5^{s_1+s_3}X_5^{s_2+s_4}$ is carried out on qubit $Q_5$, which then ends up in the state corresponding to applying $U_{\rm rot}(\xi,\eta,\zeta)$ on the input state.

\begin{figure*}
\begin{center}
\includegraphics[width=\linewidth]{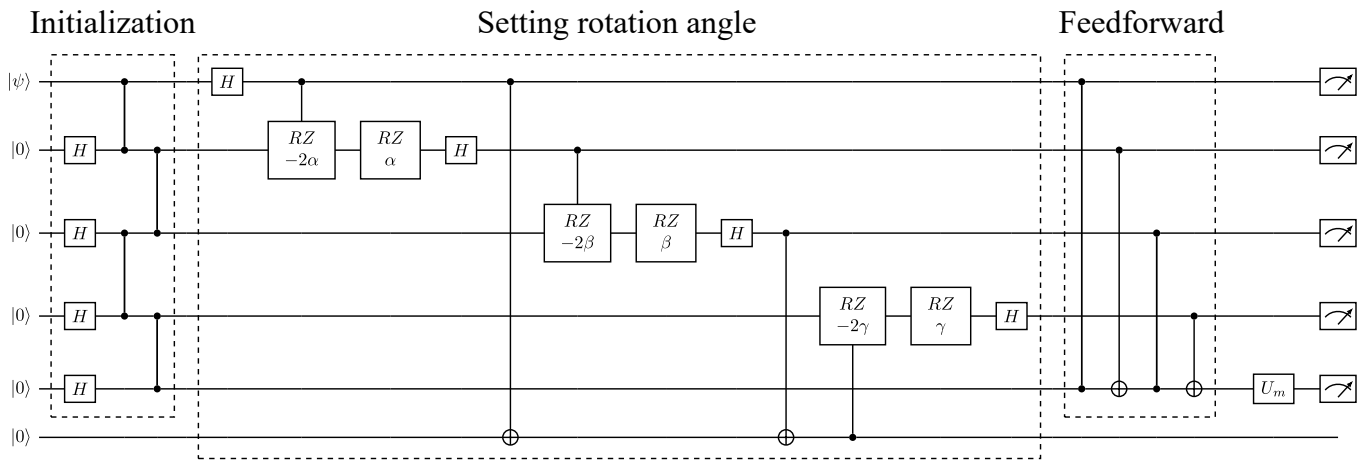}
\end{center}
\caption{The circuit for the delayed-choice one-way arbitrary rotation gate. The input and output states are encoded in qubits $Q_1$ and $Q_5$, respectively. The gate $U_m$ before measuring qubit $Q_5$ represents a rotation operation that decides the measurement basis for $Q_5$.}
\label{Rcirc}
\end{figure*}

We note that the measurement direction of qubit $Q_4$ is determined by the both the measurement results $s_1$ and $s_3$, which makes it impossible to apply delayed choice directly. In order to solve this, we bring in another physical qubit as a register that calculates and saves the Boolean result $s_1+s_3$. This additional qubit also serves as the control qubit during the corresponding delayed choice. The delayed-choice one-way arbitrary rotation gate is shown in \figref{Rcirc}. Here, the delayed-choice one-way arbitrary rotation gate requires more SWAP operations than the Hadamard case. Due to the aforementioned analysis, we are simply trying to show the logical result of our protocol.

The results of our numerical simulations are shown in Table~\ref{tab:arbitrary}. For the noisy simulations, we used the same parameters as in the simulations in Appendix~\ref{Hadamard}.

\begin{table}
\caption{Probability distribution for the delayed-choice one-way arbitrary rotation gate $U_{\rm rot}(\xi,\eta,\zeta)=U_x(\zeta)U_z(\eta)U_x(\xi)$ on the quantum assmebly simulator. This circuit does not consider the actual qubit connectivity, but instead assumes that all the qubits can directly interact with each other as required. This table shows the simulation results with different rotation angles. During the simulation, we have considered two inputs with different rotation angles:
(a) Input state $\ket{\psi}=\ket{0}$ with rotation angles ($\frac{\pi}{2}$,$\frac{\pi}{2}$,0) and
(b) Input state $\ket{\psi}=\ket{+_y}$ with rotation angles (0,$\pi$,$\frac{\pi}{2}$).
For the simulations with noise, the fidelities of the single-qubit and two-qubit gates are set to \unit[99.8]{\%} and \unit[98]{\%}, respectively.
\label{tab:arbitrary}}

\begin{ruledtabular}
\begin{tabular}{lccc}
Input $\ket{0}$, angle $(\frac{\pi}{2},\frac{\pi}{2},0)$ & $\ket{+_x}$ & $\ket{-_x}$ \\
\colrule
Ideal & 1 & 0 \\
Noisy & 0.866 & 0.134
\end{tabular}
\end{ruledtabular}

\qquad

\begin{ruledtabular}
\begin{tabular}{lccc}
Input $\ket{+_y}$, angle $(0,\pi,\frac{\pi}{2})$ & $\ket{0}$ & $\ket{1}$ \\
\colrule
Ideal & 0 & 1 \\
Noisy & 0.137 & 0.863
\end{tabular}
\end{ruledtabular}
\end{table}

\bibliography{One-way}

\end{document}